.

# THE BUCKET BRIGADE
PRICING AND NETWORK EXTERNALITIES IN PEER-TO-PEER COMMUNICATIONS NETWORKS

*Last Update on October 22, 2001*

SAM CHANDAN[†] and CHRISTIAAN HOGENDORN[‡]

ABSTRACT – THIS PAPER analyzes the pricing of transit traffic in wireless peer-to-peer networks using the concepts of direct and indirect network externalities. We first establish that without any pricing mechanism, congestion externalities overwhelm other network effects in a wireless data network. We show that peering technology will mitigate the congestion and allow users to take advantage of more the positive network externalities. However, without pricing, the peering equilibrium breaks down just like a bucket brigade made up of free-riding agents. With pricing and perfect competition, a peering equilibrium is possible and allows many more users on the network at the same time. However, the congestion externality is still a problem, so peering organized through a club may be the best solution.

[†] Authors' Affiliation: Doctoral Candidate, The Wharton School, University of Pennsylvania, 3100 Steinberg Hall-Dietrich Hall, Philadelphia PA 19104-6372 and Master's Candidate, School of Engineering and Applied Science, University of Pennsylvania. Telephone: +1 (215) 898-9619. Electronic mail: chandan@wharton.upenn.edu.

[‡] Assistant Professor of Economics, Department of Economics, Wesleyan University, 238 Church Street, Middletown CT 06459-0007. Telephone: +1 (860) 685 2340. Electronic mail: chogendorn.web.wesleyan.edu.

# I. INTRODUCTION

While peer-to-peer (P2P) networking has been made famous by Napster and similar distributed file sharing systems, one of its most promising commercial applications is in wireless communications networks. As in the case of conventional wireless networks, there are base stations and users with mobile devices in wireless P2P networks. But unlike their conventional counterparts, the mobile devices in a P2P network are capable of communicating directly with one another as well as with the network's base stations. Subject to some maximum distance limitation, a mobile device can initiate a direct link with another destination device. Alternatively, the originating device can complete the link through a peering connection, where the resources of several intermediate devices are used to complete the connection. The process is analogous to a bucket brigade fighting a fire: the message is passed from one user to another until it reaches the destination. This peering capability has tremendous technological and cost advantages over conventional wireless networks because spectrum can be reused very efficiently and the number of users supported by a single base station can be increased greatly.

In this paper, we analyze the pricing of transit traffic in a wireless P2P network using the concepts of direct and indirect network externalities. When an additional user joins a P2P communications network, the value of the network increases for two reasons: (1) the direct network effect: as the membership of the network grows, the number of potential connections increases. Existing users can communicate with the new user over the P2P network rather than communicating by some costly interconnection with other networks, and (2) the indirect network effect: the existing users will find that the network has more capacity and higher service quality because the new user adds another intermediate station to the network.

The effect of membership growth is not unambiguously positive. The immediate problem is one that characterized P2P networks generally: because P2P networks are decentralized, each user's mobile device must dedicate some of its bandwidth and battery power (or whatever the resource in question might be) to facilitating the transit traffic of other users. But each user has an incentive to make use of others' bandwidth without actually carrying any transit traffic him- or herself, wireless P2P networks, like the popular Internet-based P2P networks, demonstrate the free-rider problems associated with public goods. This property of P2P networks has been documented by Adar and Huberman (2000) in their study of Gnutella's music file sharing network. The authors speculate that the absence of monitoring will result in free riding as a P2P network grows larger, with users consuming more resources and sharing fewer resources. In support of their argument, the authors observe Gnutella's network traffic over a 24 hour period, finding that 70% of users do not share files and that 50% of file requests are returned to the top 1% of sharing hosts. In addressing the "tragedy of the *digital* commons" technical and market solutions are proposed. The latter, the creation of a market system whereby public goods are effectively made private, is analogous to the wireless P2P network presented in our model, where prices are used to enable a market that otherwise fails to materialize because of a divergence in objectives between aspiring free riders and the providers of resources.

The second reason to expect that growth in the network is not unequivocally positive is a problem characteristic of all wireless networks: positive network effects are confounded by a congestion externality. While an originating device incurs the costs of establishing a connection with its final or peering destination, it does not bear the costs of the spectrum interference generated by its connection except through some market mechanism. This problem is analogous to the bandwidth scarcity that characterizes conventional cellular telephone networks.

We assume throughout the paper that the network itself is enabled by competitive hardware manufacturers who do not attempt to internalize the network effects. In our base case, we examine the outcomes in a market where no prices and control mechanisms are available to moderate the behavior of the agents. We then model two scenarios explicitly: (1) a decentralized market in which a competitive market price for transit traffic is established, (2) a club market in which users form a "club" that controls entry to the network. We first establish that in the decentralized, competitive market, true network externalities occur, related to both positive network effects and pollution. We next determine a competitive price that enables a peering market but still leads to sub-optimal outcomes. Finally, we examine a club market. We show that a club can internalize the network externalities. We conclude with some implications for regulatory policy which suggest that certain types of entry controls may actually be welfare-enhancing in the P2P environment.

*Peer-to-Peer Networking*

The dominant computer network architecture of the last two decades has been client-server. In this paradigm, individual computers are connected to one another indirectly, via centralized servers. While functions and processes that are essentially local are performed and executed locally using an individual computer's resources, servers provide shared resources and services such as centralized file sharing, peripheral sharing, access to other networks and, in the case of specialized transaction or database functions, server-side processing. Emerging P2P networks upset this paradigm, primarily by facilitating direct connections – usually virtual and sometimes physical – between individual computers. While no formal definition exists for P2P networking, the essential feature of these networks is that each node has a roughly equivalent set of capabilities and responsibilities. Resources and control are decentralized. This contrasts with the client-server architecture where nodes are specialized, the server nodes' explicit responsibility being to serve end-users in some way.

In practice, fully decentralized systems are difficult to implement. Hybrid systems, where some features and functions are centralized and others decentralized, are more often observed in practice than the 'pure' P2P model. While P2P has been popularized by Napster and similar file sharing applications, such decentralized models of control in computer networks are not new. Nor is P2P limited to file caching and sharing applications. The earliest applications of the nascent Internet exhibited some decentralized characteristics. For example, Telnet and File Transfer Protocol (FTP) allowed any computer on the network to assume a client role in connecting to other computers. The remote computer then acts a as server for remote processing or file upload and download. Microsoft Windows has included a peering network protocol for over a decade, since the introduction of Windows for Workgroups. Windows peering obviated the need for a client-server local area network by allowing a small group of computers, connected by standard Ethernet, to access resources such as printers and hard drives with local connections to specific computers on the network. Instant messaging (IM) is a more recent P2P application. IM allows text messages to be sent directly between users. However, a centralized name and presence database (NPD) is required for coordinating and disseminating online state information. Indeed, Napster uses a similar hybrid model. Bandwidth intensive file sharing takes place between clients at the periphery of the network while the file directory is centralized on Napster's servers.

Because P2P networks share resources in a decentralized control model, these networks are typically characterized by a public goods problem. Formally, the public goods problem arises when a good is nonrivalrous or nonexcludable in use. A good is rivalrous when consumption by one individual reduces the amount available to consume by other individuals. A good is excludable when individuals can be excluded from consuming it, usually because one individual or group of individuals has exclusive control of the good.  A P2P network might be characterized by a *pure* public goods problem or by a *free* public goods problem.  The pure public good is nonrivalrous and nonexcludable.  In the absence of some institutional mechanism for exclusion, the files of a music sharing application are pure public good.  The free rider in this case is the user who habitually downloads music from his peers without also making music available to others.  Even in the absence of free riding, the bandwidth required to transfer the files is a free public good, also referred to as a common property resource.  While nonexcludable, the bandwidth-intensive file transfer is rivalrous in that it reduces the resources available to other users of the network.

The bandwidth scarcity of the wired network has an imperfect analogue in the spectrum scarcity problem of its wireless counterpart.  And while conventional wireless networks already contest with the problem of spectrum scarcity, the free public goods property of P2P wireless networks is exacerbated by the shared resources contributed by each device.  The common property characteristics of contributed resources will result in the wireless P2P network conforming to the general outcome observed for free public goods: nonexcludability will lead to overconsumption of spectrum and others' peering resources while contemporaneously reducing the incentives for each individual to make costly investments in resources that will then be shared with greedy peers.  Open access leads to a clear efficiency loss from overconsumption and underprovision.  Some structural or institutional mechanism must be introduced to correct the inefficiency.

*Technology of Wireless Communications*

The public goods problem is one of two market failures that will impede efficient outcomes in P2P wireless networks.  The externality feature, discussed in the next section, is not specific to wireless communication.  Rather, it is a general feature of networks that the value of membership is increasing in the number of members.  Conversely, the specific nature of the public goods problem does result from the technology of wireless communications.  A basic understanding of this technology is necessary to the development of the model.

While cellular telephones and other mobile communications devices bear a clear functional resemblance to two-way wireline communications devices, the technology of wireless communications is closer to that of radio.  Both wireless communications and radio use wave propagation on the electromagnetic spectrum to transmit voice and, more recently, data.  These electromagnetic carrier waves are characterized by their frequency and their amplitude, where frequency is defined by the number of cycles per second, measured from one wave peak to the next. While early wireless systems used a fixed frequency and amplitude modulation (AM), the susceptibility of AM signals to ambient environmental conditions requires a more robust signal. In modern wireless system, voice is superimposed on a carrier wave of constant amplitude; the frequency is modulated to capture variations in sound.

The first wireless systems used a broadcast model where a specific frequency band was allocated for use in a defined geographic area of service. Within this area, the specific frequency would be used for the transmission and reception of radio signals. While frequencies could be subdivided into channels, the finite frequency band assignment and the broadcast nature of the system limited the number of simultaneous, full duplex connections to one-half the number of channels.

Cellular systems are distinguished from broadcast by frequency reuse and connection handoff. Cellular replaces the broadcast model's single high-powered radio transmitter with many low-powered transmitters. While the former services a large area, each of the smaller transmitters serve a smaller "cell". In the case of frequency reuse, the finite allocation of frequencies is divided between cells so that no single frequency band is reused in an adjacent cell. While a weak signal on the same frequency may reach a nearby cell where that frequency is being reused, the property of FM capture dictates that, given the reception of two signals on the same frequency, the transmission antenna will focus on the stronger of the two signals. The weaker signal is suppressed at the transmission antenna. Handoff is a technology for allowing mobile devices to move between these cells. As a device moves further from the center of the its current cell and towards an adjacent cell, the relative signal strengths fall and rise. When signal strength from the adjacent cell is stronger than in the current cell, a handoff occurs.

The introduction of peering to a cellular network would enhance the network's capacity for frequency reuse by reducing the mean distance between any two devices in a connection. Nonetheless, the peering network would inherit the properties of the electromagnetic spectrum and of the cellular design:
- Bandwidth scarcity – unlike wireline communications, where fiber optic and other technologies can be deployed to address bandwidth scarcity, the electromagnetic spectrum is limited. The general principles described above are formalized in Shannon's Law and an extension by Webb. Shannon's Law shows the maximum number of bits that can be transmitted per second per hertz of spectrum, where SNR is the signal-to-noise ratio at the receiver:

$$C = \log_2(1+\text{SNR})$$

Substituting the signal-to-interference ratio and optimizing the system, the maximum number of channels per cell is given by the following, where $B_t$ is the available bandwidth, $R_b$ is the user bit rate and $a$ is factor, approaching 1, related to the system's proximity to Shannon's limit:

$$m = 1.42 \cdot a \cdot \frac{B_t}{R_b}$$

Another characteristic of a wireless network relates to path loss. Path loss formalizes the relationship between distance, power, frequency and signal strength and is given by:

$$L = \frac{P_r}{P_t} = \frac{K}{f^2 \cdot d^n}$$

where $P_r$ represents the received power, $P_t$ represents the transmitted power, $f$ represents the frequency of the transmitted signal, $K$ is a constant that represents the average path loss at a reference distance from the transmitter and $n$ represents the proportional decrease in signal strength relative to distance.

From the path loss equation and other basic characteristics of electromagnetic waves, we can infer the remaining physical properties that should be accounted for in the P2P wireless model:

- Coverage – signal strength diminishes proportionally to the square of the distance from the transmitter. The theoretical relationship is $d^2$ but is closer to $d^4$ in practice.
- Signal error – Signals are subject to attenuation, which is aggravated by transmission through the physical environment.
- Power – To effectively transmit signal over longer distances requires more power.

*Economic Properties of Communications Networks*

While the discussion of wireless technology introduced the free public goods properties that are particular to this form of communication, the network externalities phenomenon is a more general feature of communications networks. Many goods and services, often organized in networks and often not (Economides & Salop 1992), exhibit positive consumption externalities, commonly referred to as network externalities or network effects. It is important that network effects should be distinguished from network externalities: network effects refer to the case where "the net value of an action is affected by the number of agents taking equivalent actions" (Liebowitz & Margolis 1994). In the second case of network externalities, a true market failure occurs as participants fail to internalize the effects: "the equilibrium exhibits unexploited gains from trade regarding network participation". Our model of the wireless P2P network is characterized by network externalities; the network is smaller than the social optimum. Where network externalities are present, the value of consuming a good or service increases in the number of other consumers of that good or service. Formally, network externality is an individual user's change in utility that derives from consumption of a good or an action as the number of other consumers of that same good or number of agents taking equivalent action increases (Katz and Shapiro 1985). While much attention is given to the positive form, negative externalities can be present, as well. Conduit congestion, whether the conduit is a wire or the electromagnetic spectrum, is an example of a negative network externality often referred to simply as a congestion externality.

Goods which exhibit network externalities can be separated into two distinct types (Economides and Flyer 1997). The first type of good, termed a pure network good, derives all of its value from the network externality. The second type of good has value even in the absence of a network. The *autarky* value is the value of this good in the absence of other users, while the *synchronization* value embodies the externality. In the case of the pure network good, the autarky value is zero.

The telecommunications industry is a standard example of an industry characterized by a strong pure network externality. The externality may exist because of the expectations of agents, coordination among agents, or because of the complementary nature of the good being consumed. Again, the telephone network provides an illustrative example of the expectations and complementarities foundations of the externality. If only one person in a market owns a telephone, he has no one else to call. The benefit he derives from having the telephone, or 'node', is zero. If one other person in the market purchases a telephone and the two phones are connected by a 'link' such as a twisted-pair, each of the two owners has one other person that they can call. This one connection is the numeraire good, so the aggregate value of the network is now two. Specifically, $n_i$ calling $n_j$ is considered as distinct from $n_j$ calling $n_i$ in this two-way network. If a third person purchases a telephone, he can make calls to the two existing owners. His benefit is two. However, the number of potential connections in the network has increased from two to six. This third member of the network internalizes and makes decisions based on his private benefit of two. He does not consider the increase in value to the other members of the network – this is the externality. As the number of telephones in the network increases, the number of potential connections continues to increase. For *n* telephones, the number of potential connections is $n \cdot (n-1)$. Each additional telephone increases the number of potential connections by $2 \cdot n$.

Network externalities can be direct, as in the case of the telephone, and indirect. The former are described as direct because the addition of a new user directly increases the number of potential connections, hence the value of the network. In many two-way networks and most one-way networks, there is an indirect or complementary externality, as well. The indirect form of the network externality is present in networks when consumers demand composite goods (Church and Gandal 1992). Individual consumer utility and network size are related circuitously, through the variety of complementary goods. Returning to the example of the telephone network, we now consider a composite network where users demand telephones and a complementary good, answering machines. As each new user joins the network, there is the direct effect described above and an indirect effect that results from the increase in demand for answering machines. If higher demand for the complement increases the variety of the complement, for example, the indirect effect is positive.

## II. THE MODEL

Our model of the P2P wireless network is presented in two stages. In the first stage, a user *i* on the network decides to make a call to some other user *m*. In the second stage, *i* completes the connection to *m* subject to the constraints of the particular peering environment. We first describe the basis properties of the market and technology. We then describe a no-peering base case, peering under perfect competition, and peering with clubs.

*Geography and Connections*

We take a stylized geography to analyze the problem. Let there be an infinite plane on which are located nodes. Let the nodes be uniformly distributed over the plane with the density in any 1x1 square equal to $n^2$. For simplicity, let the nodes be arranged in equally-spaced rows and columns.

We assume that communication between nodes is bi-directional, i.e. that the cost of establishing and maintaining a connection is symmetric between any pair of nodes, but is costly in proportion to the distance between those nodes. The distance between two nodes $i$ and $m$ is denoted by $d_{im}$.

The cost of connection between two nodes is denoted $c(d_{im})$. The cost is convex in distance, $c' > 0$ and $c'' > 0$. This reflects the increase in power needed to send signals over longer distances.

A connection between two nodes can be direct. In this case, $i$ incurs the full cost of establishing the connection to $m$ over a distance $d_{im}$.

Alternatively, the connection can be peering. In the case of a full peering connection, $i$ establishes a connection with the nearest node (its nearest neighbor) in the direction of the destination node. This nearest neighbor in turn incurs the cost of establishing a connection with its nearest neighbor and so on, until node $m$ is reached and the connection is completed. Given a path between two nodes on the network $i$ and $m$, the length of the path is defined as the sum of the distances between the initial, intermediate, and destination nodes: $d_{ij} + d_{jk} + ... + d_{lm}$. The number of intermediate nodes between $i$ and $m$ is denoted $I_{im}$. For simplicity, we will let

$$I_{im} = I(d_{im}) = nd_{im} - 2$$

This is an approximation which is only exactly correct when communication takes place along rows, columns, or diagonals. The number of transmissions required for the peering connection is $1+I(d_{im})$, i.e. one transmission from the originator plus one from each of the intermediate nodes.

*Bandwidth and Congestion Externalities*

Unlike a voice telephone network, in which each call takes up a uniform amount of bandwidth, P2P wireless networks will primarily serve high-bandwidth, packet-switched applications. As such, we expect that each connection will potentially use a large proportion of the available bandwidth. Such high usage of bandwidth is not actually *necessary* to make the connection, but it increases the quality of service. Thus, we model each connection as being feasible at all times, but the originator's utility is reduced by other people using the network at the same time.

Suppose that node *i* transmits to node *m*, either directly or as one link in a peering connection. The signal produced by *i* takes up spectrum that could be used by all other nodes in a circle of radius $d_{im}$ around *i*. The number of nodes within this circle of radius $d_{im}$ is:

$$N(d_{im}) = \pi d_{im}^2 n^2 - 1$$

Each of the nodes in the circle suffers a pollution cost *w* due to the inability to use this spectrum. This cost reflects the opportunity cost of not being able to send and receive packets as quickly as would otherwise be possible. Thus, the total pollution cost of a signal over distance $d_{im}$ is given by $wN(d_{im})$. Note that $N()$ and hence $wN()$ is increasing, convex with distance.

*Decision to Make a Connection: the Direct Network Effect*

The universe of nodes that node *i* might connect with includes only those nodes within a cell of some maximum radius. Beyond this radius, mobile devices do not have sufficient power to overcome fatal path loss. As such, devices will complete longer distance connections through a base station, which we take as exogenous to the model. Thus, we consider only nodes in a circle of radius $d_{max}$. There are $N = N(d_{nax})$ such nodes.

A node decides to make a connection according to the following process:

The node looks at every possible peer that it could connect with. For each peer, there is a probability $(1 - z)$ that the node would get positive utility from a connection. If there are $N$ peers, the probability that a node wants to connect with at least one of them is $P(N) = 1 - z^N$. Let the set of peers that node *i* wants to connect to be denoted $\mathbf{N}_i$. We will assume that for one randomly selected $m \in \mathbf{N}_i$ the utility from a connection is $v>0$, and from all other $k \in \mathbf{N}_i$ the utility is $u>0$. We further assume that $v - u > c(d_{max})$, which implies that node *i* is always willing to pay to connect to *m* and that it is never worthwhile to call a closer node in order to save money on the cost of the connection.

We choose these assumptions for two reasons: (1) We do not believe the differences in battery power used for shorter- versus longer-distance connections are likely to be large relative to the differences in valuation of connecting with the most desired party versus the next-most desired party. (2) Although P2P networking will reduce the cost of longer-distance connections, we do not believe that it is plausible that it will significantly affect the distances over which consumers choose to make connections. Of course, this conclusion would be wrong if P2P included substantial caching or mirroring of material at each node, as with Napster. The type of P2P network we are modeling here does not include such storage possibilities.

**Base Case: No Monetary Transfers Among Nodes**

We begin with a very simple setting in which there is no technology to transfer money among the nodes. We will show that in this case, peering may be desirable, but it is never an equilibrium outcome because of free-rider problems.

During the instant of time that we model, we assume that each node has one opportunity to make one connection. Since the nodes are mobile, we will assume that a node is in the center of the circle when it makes a connection. Since the peers are all arrayed in the circle and connections are made randomly, there is a probability distribution over the distance of a connection, denoted $f(d)$.

We derive $f(d)$ by noting that the probability of a randomly selected destination node among $N(d_{nax})$ total nodes being within radius $d$ is

$$F(d) = \frac{N(d)}{N(d_{max})} = \frac{pd^2 n^2 - 1}{pd_{max}^2 n^2 - 1} \approx \frac{d^2}{d_{max}^2}$$

where we take the approximation because we assume that the network is sufficiently large. In this case, $f(d)=2d/d_{max}$. The important property to node is that $f()$ does not vary in $n$, assuming $n$ is large.

The utility of the originating node from making a connection is

$U(\text{ORIG, NOPEERING}) = v - c(d_{im})$

The expected utility of the originating node is the probability of wanting to make a connection times the expected gain from the connection:

$$EU(\text{ORIG, NOPEERING}) = P(\overline{N})\int_0^{d_{max}} (v - c(x))f(x)dx$$

This shows the direct network externalities at work. As the number of users of the network goes up, the probability of wanting to connect with one of them, $P(N)$, goes up. The effect is concave in the density of users: $dP/dn > 0$, $d^2P/dn^2 < 0$.

In addition to each node having a probability of being an originator, there is a corresponding chance of being an "outsider" subject to pollution from other nodes making connections. The utility of being an outsider is

$$U(\text{OUTSIDE, NOPEERING}) = -w$$

The probability of being an outsider is the probability of each other node actually making a connection times the probability of being within the distance of the that node's transmission. All of these are independent random events, so

$$EU(\text{OUTSIDE, NOPEERING}) = -w\overline{N}P(\overline{N})\int_0^{d_{max}} \frac{N(x)}{\overline{N}} f(x)dx$$

$$= -w \begin{pmatrix} \text{Expected} \\ \text{connections} \\ \text{made by} \\ \text{other nodes} \end{pmatrix} \begin{pmatrix} \text{Probability of being within} \\ \text{range of another node's} \\ \text{transmission} \end{pmatrix}$$

This shows the congestion externalities as work. As the number of users of the network goes up, expected number of total connections is increasing convex, while the probability of being within range of any one connection remains constant.

The total utility of a node is $EU(\text{ORIG, NOPEERING}) + EU(\text{OUTSIDE, NOPEERING})$. This total utility is eventually decreasing in $n$, the density of nodes in the network. This is because the direct network externality is increasing concave in $n$, while the congestion externality is increasing convex in $n$. If there is free entry and no alternative network, nodes will continue to join the network until $EU(\text{ORIG, NOPEERING}) + EU(\text{OUTSIDE, NOPEERING}) = 0$.

*Peering in the Base Case*

Now consider what would happen if the peering technology were available, but no monetary transfers were possible. If, nevertheless, all nodes were willing to peer, then a connection from $i$ to $m$ would be completed as follows. The distance between $i$ and $m$ remains $d_{im}$, and there are $I(d_{im})$ intermediate nodes between $i$ and $m$. The distance between each of the intermediate nodes is given by:

$$D(d_{im}) = \frac{d_{im}}{1+I(d_{im})} = \frac{d_{im}}{nd_{im}-1}$$

We will consider the peering decision in three steps: will the originating node want to peer, will the intermediate nodes want to peer, and will the outside nodes want peering.

The originating node $i$ faces a decision to connect directly to $m$ or to peer. If it peers, it need only send a transmission to the first intermediate node. Thus, $i$'s objective function is:

$$\max\ [v - c(d_{im}), v - c(D(d_{im}))]$$

The origin node maximizes the difference between the value of the connection and the cost of implementing the connection. The cost is framed in terms of either the direct connection to the destination or a peering connection, where the origin only incurs the cost of reaching its nearest neighbor. The following results clearly follow from this simple maximization:

① if $m$ is $i$'s nearest neighbor, the connection is made directly;
② otherwise, $i$ desires a peering connection.

The expected utility in this case is

$$EU(\text{ORIG, PEERING, NOTRANS}) = P(\overline{N}) \int_0^{d_{\max}} (v - c(D(x))) f(x) dx$$

We next consider the intermediate nodes. Suppose that intermediate node $k$ does offer peering. Since it receives an incoming signal, it suffers pollution cost $w$. Since it also needs to relay this signal, it incurs a direct cost given by $c(D(d_{im}))$ to reach the next node $l$, provided that $l$ also offers peering. If an intermediate node does not offer peering but the adjacent nodes do, it just suffers a pollution cost *w*.

The objective function of the intermediate node *k* follows:

$$\max [-w - c(D(d_{im})), -w]$$

Clearly, then, peering will not occur without some type of incentive structure.

Now consider the pollution costs to the outsider nodes. With full peering, each of the $I(d_{im})+1$ originator plus intermediate nodes broadcasts a signal over a distance of $D(d_{im})$. The expected utility from being an outsider is the probability of a call being made times the probability of being within range of one of the peering signals associated with the call:

$$EU(\text{OUTSIDE, PEERING, NOTRANS}) = -w\overline{N}P(\overline{N})\int_0^{d_{\max}} \frac{(I(x)+1)N(D(x))}{\overline{N}} f(x)dx$$

$$= -w \begin{pmatrix} \text{Expected} \\ \text{connections} \\ \text{made by} \\ \text{other nodes} \end{pmatrix} \begin{pmatrix} \text{Probability of being within} \\ \text{range of the peering transmissions} \\ \text{made by each other node and} \\ \text{its peers} \end{pmatrix}$$

As before, the expected total number of connections is increasing convex in *n*. But with peering, the probability of being within the range of any one connection is actually decreasing in *n*. The net result in the stylized geography is that with peering, the congestion externality increases linearly in *n*, a large improvement over the convex increase without peering.

*Desirability of Peering*

Without peering, the total costs to society of a particular connection are given by $c(d_{im}) + wN(d_{im})$. With peering, the total costs are given by $(I(d_{im})+1) \cdot c(D(d_{im})) + (I(d_{im})+1)wN(D(d_{im}))$. Because $c()$ and $N()$ are both convex, the total costs are always lower under peering.

Now suppose all the nodes are peering, even though it is not in their self-interest to do so. We want to examine the value added by one of these nodes due to its willingness to peer. If node *j* does not peer, then node *i* will have to skip over *j* and transmit directly to *k*. The total costs in this case are $(I(d_{im})-1)c(D(d_{im})) + c(2 D(d_{im})) + (I(d_{im})-1)wN(D(d_{im})) + wN(2 D(d_{im}))$. Thus the value added to the system by node $j$ (net of the cost to $j$) is:

$$VA = -2c(D(d_{im})) + c(2 D(d_{im})) - 2wN(D(d_{im})) + wN(2 D(d_{im}))$$

Because $c()$ and $N()$ are convex, $VA$ is always positive. This means that the contribution to total welfare that $j$ makes by peering is greater than the cost $j$ incurs by peering. We might then want to consider a way of compensating node $j$ to convince it to peer.

As a way of examining whether such a scheme is possible, consider that the value of peering to the originating node $i$ is $c(d_{im}) - c(D(d_{im}))$

This savings would be enough to compensate each of the intermediate nodes for the cost of peering. Node $i$ would still be left with a net gain even after the compensation, plus there would be a gain to society from the reduced pollution.

**Peering with Perfect Competition**

We have shown that node $i$ will always want to establish the connection through peering, except where the destination node is $i$'s nearest neighbor. In the absence of prices, however, there is no incentive for intermediate nodes to offer peering. Node $i$ must compensate each intermediate node through which it establishes a peering connection to at least offset the additional costs the intermediate node bears. It does this by paying some price $p$ to each intermediate node:

$$\max\left[v - c(d_{im}),\ v - c(D(d_{im})) - I(d_{im})p\right]$$

The utility and expected utility of the connection now includes this price:

$$U(\text{ORIG, PEERING, PERFCOMP}) = v - c(D(d_{im})) - I(d_{im})p$$
$$EU(\text{ORIG, PEERING, PERFCOMP}) = P(\overline{N})\int_0^{d_{max}} (v - c(D(x)) - I(x)p)f(x)dx$$

The intermediate node $j$ includes the transfer in its maximization function:

$$\max[-w - c(D(d_{im})) + p, -w]$$

$p_{i,i+1} \geq c(D(d_{im}))$ The intermediate node will offer peering and the originator will pay for it when the price is bounded by the costs of peering and a direct connection:

$$c(D(d_{im})) \leq p_{i,i+1} \leq c(d_{im})$$

We assume that the intermediate node does not know the number of other nodes or the positions of other nodes that might offer peering service for this connection. In this imperfect information scenario, the intermediate node acts in a competitive manner. It is clear that in a competitive setting, supply will be priced at marginal cost. Hence, the competitive price offered by the intermediate node is:

$$p_{i,i+1} = c(D(d_{im}))$$

And this means that the utility of the intermediate node is just $-w$. Then we can derive the expected utility for the originator, intermediaries, and outsiders as we did before:

$$EU(\text{ORIG, PEERPING, PERFCOMP}) = P(\overline{N}) \int_0^{d_{max}} (v - (I(x)+1)c(D(x)))f(x)dx$$

$$EU(\text{INT, PEERING, PERFCOMP}) = -w\overline{N}P(\overline{N}) \int_0^{d_{max}} \frac{I(x)}{\overline{N}} f(x)dx$$

$$EU(\text{OUTSIDE, PEERING, PERFCOMP}) = -w\overline{N}P(\overline{N}) \int_0^{d_{max}} \frac{(I(x)+1)(N(D(x))-1)}{\overline{N}} f(x)dx$$

The top line expresses the direct network externality, which is increasing concave in $n$. It also expresses the indirect network externality that costs are lower as $n$ increases, the benefit of this is also increasing concave in $n$. The middle line expresses the indifference of the peering nodes under perfect competition. The middle and bottom lines express the congestion externality, which is increasing linearly in $n$. Again, the sum of these expected utilities eventually decreases in $n$, but more slowly the so the free entry number of nodes, where the sum of all three lines equals zero, is much larger than in the no-peering network.

**Peering with Clubs**

As with all congestible resources, entry of additional users until the marginal user receives zero utility does not maximize the utility possible from the network. Two alternatives that might be thought to increase utility are adjustments to the price of peering and the number of entrants on the network. We first discuss the latter alternative, a direct control on entry to the network. Then we consider directions for future research related to the price of peering which are beyond the scope of the present model.

*Clubs with Entry Controls*

In conventional wireless networks without p2p, such as existing cellular and PCS telephone networks, there is a form of "entry control" in the sense that users must pay a positive price for network access and for minutes used. These prices are based on mobile operators' marginal cost of base station usage for each additional user and also on these operators' market power in many areas. In the pure p2p network we are modeling, no such costs or market power exist. Thus, under perfect competition, entry continues to occur until the benefits of joining the network are zero (or perhaps until the benefits are equal to the cost of the handset needed to use the network).

However, this number of users is obviously too large because all users are indifferent between joining the network rather than receiving a benefit. A club could institute an entry fee, or equivalently a limitation on bandwidth used, that would help to reduce the congestion externality. The club could be run for profit, in which case it could charge an entry fee and keep it, or it could be a voluntary association in which case some number of "slots" would be rationed off and thereafter entry of a new user could only occur after the resignation of an existing user.

In either case, the market would behave as in the perfect competition case once the number of entrants was set, so the club would seek to maximize

$$EU(ORIG,PEERING,PERFCOMP) + EU(INT,PEERING,PERFCOMP)$$
$$+ EU(OUTSIDE,PEERING,PERFCOMP)$$

This maximization results in a reduction in the number of users relative to the free-entry case under perfect competition. The users who remain receive a benefit from membership, or they may transfer this benefit to a profit-making club owner through a membership fee.

*Clubs with Price Controls*

Adjusting the price of a peering connection to reflect the pollution externality follows the usual first-best solution to externality problems in economics. We have vitiated its affect here with our modeling assumption that the probability of a connection, $P(N)$, is independent of price. This means that raising the price of peering cannot reduce the number of connections. Therefore, the only effect of an increase in the price of peering above the perfectly competitive level is to transfer money from the originator to the intermediaries. Since each node has the same probability of finding itself in these two roles, the net effect of the price change is zero.

Now consider what would happen if $P()$ were somewhat responsive to price (i.e. if demand were inelastic rather than perfectly inelastic). Then an increase in the price of peering would indeed reduce demand for connections and hence reduce congestion. But if demand were inelastic, a rather large price increase would be required to offset the full amount of the congestion effect. However, if the price of peering were to rise above $c(2D(d_{im}))$ for any given connection, there would be an incentive for coalitions of nodes to form in order to create "leapfrog" peering paterns where some intermediary nodes are skipped. While such patterns would bring profits to those who created them, they would also involve longer transmission distances for the intermediaries and would therefore cause greater suggestions. For this reason, we believe a fruitful avenue for further research is to show that the usual Pigouvian solution to the congestion externality will actually cause *more* congestion by disrupting the peering equilibrium.

### III. Conclusion

Packet-based wireless services are a complex networked good that contains three important network externalities. First, there is a direct network externality because more users will increase the value of the network by giving more destination options. Second, there is an indirect network externality that occurs when peering is used because the peering system functions at lower cost when there are more users of the network. Third, there is a congestion externality because additional users' transmission will reduce the bandwidth available to other users. All three are true externalities in the sense that they are not internalized by market forces. The first two will make free-entry networks too small, while the third will make them too large. We have shown that in the simple geography, the congestion externalities will always come to dominate and the network will be too large. This assumes that no additional technology is applied to mitigate congestions.

One likely candidate for reducing congestion is peering, which allows messages to be relayed as if by a bucket brigade. In effect, peering technology is a network externality "enabler." That is, it allows users to enjoy the benefits of positive network externalities without causing too much congestion. In so doing, it actually creates more positive network externalities since peering itself works better when there are more users.

However, the technology is not perfect, because it mitigates but does not eliminate the congestion effect. For this reason, a club system may be more efficient. However, it may also appear to be an effort to control what would otherwise be a competitive market. Regulatory scrutiny of such arrangements should be tempered by the congestion aspect of the problem. Opportunities for the entry of entirely new clubs are probably much more important than the behavior of existing clubs in this setting.

<div style="text-align:center">**REFERENCES**</div>

___